# Directly visualizing the momentum forbidden dark excitons and their dynamics in atomically thin semiconductors

Julien Madéo* (1), Michael K. L. Man* (1), Chakradhar Sahoo (1), Marshall Campbell (2), Vivek Pareek (1), E Laine Wong (1), Abdullah Al Mahboob (1), Nicholas S. Chan (1), Arka Karmakar (1), Bala Murali Krishna Mariserla (1,3), Xiaoqin Li (2), Tony F. Heinz (4,5), Ting Cao (4,6), and Keshav M. Dani (1)

(1) Femtosecond Spectroscopy Unit, Okinawa Institute of Science and Technology, 1919-1 Tancha, Onna-son, Okinawa, Japan 904-0495

(2) Physics Department, Center for Complex Quantum System, the University of Texas at Austin, Austin, TX, U.S.A, 78712

(3) Department of Physics, Indian Institute of Technology, Jodhpur, Rajasthan, India-342037

(4) Department of Applied Physics, Stanford University, 348 Via Pueblo Mall, Stanford, California 94305, U.S.A.

(5) SLAC National Accelerator Laboratory, Menlo Park, California 94720, USA

(6) Department of Materials Sciences and Engineering, University of Washington, Seattle, WA, U.S.A, 98195

* equal authors

Author e-mail address: KMDani@oist.jp

**Abstract**

Resolving the momentum degree of freedom of excitons – electron-hole pairs bound by the Coulomb attraction in a photoexcited semiconductor, has remained a largely elusive goal for decades. In atomically thin semiconductors, such a capability could probe the momentum forbidden dark excitons, which critically impact proposed opto-electronic technologies, but are not directly accessible via optical techniques. Here, we probe the momentum-state of excitons in a $WSe_2$ monolayer by photoemitting their constituent electrons, and resolving them in time, momentum and energy. We obtain a direct visual of the momentum forbidden dark excitons, and study their properties, including their near-degeneracy with bright excitons and their formation pathways in the energy-momentum landscape. These dark excitons dominate the excited state distribution – a surprising finding that highlights their importance in atomically thin semiconductors.

**Main text**

The discovery of two-dimensional (2D) semiconductors launched exciting opportunities in exploring excited-state physics and opto-electronic technologies [1,2,3], driven in-part by the existence of robust, few-particle excitonic states. As a prototypical 2D semiconductor in the transition metal dichalcogenide (TMD) family, $WSe_2$ monolayers exhibits a band structure that hosts 2 degenerate valence-band maxima, but 8 nearly degenerate conduction-band minima in the hexagonal Brillouin zone (BZ) [4,5]. At the K and K' valleys, the conduction- and valence-band energies are both at local extrema, giving rise to 2 direct bandgap transitions and the bright excitons, (denoted as K-K excitons) (Fig. 1a and 1b). These excitons are behind the strong light absorption and photoluminescence in the $WSe_2$ monolayer [6,7], and have been extensively investigated in various optical spectroscopy experiments.

Few experiments, however, are capable of probing the momentum-forbidden dark excitons in monolayer TMDs, consisting of an electron and a hole residing at different valleys [8]. Such dark excitons may interact with bright excitons, serve as the preferred carriers of information and energy, or form collective states such as exciton liquids and condensates [9,10]. As such, determining the properties and controlling the population of the dark excitons, as well as their interactions with the bright excitons, is the key to a complete understanding of the underlying physics and developing future technologies. Because of the six other conduction-band minima at the $\Sigma$ valleys in $WSe_2$ monolayers, dark excitons may form with an electron in the $\Sigma$-valley and a hole in the K- (or K'-) valley (Fig. 1a) [5,11]. The crystal-momentum mismatch between the electrons and holes make them inaccessible in the first-order optical processes such as absorption and photoluminescence [12,13].

Momentum resolved studies of excitons have been a long-standing goal [14-21]. Such studies would provide the resolution to directly access the recently sought-after dark excitons in monolayer TMDs [22, 23]. In general, ARPES based techniques have been one of the most successful in providing momentum information [24]. For example, ARPES techniques have successfully probed *free carriers* in bulk TMDs [25-28] and specially prepared monolayers [29-31]. However, observing strongly bound, few-particle excitonic states is not straightforward even conceptually, as discussed in a number of recent theoretical studies [17-21]. Experimentally, serious challenges include the need for ultrafast high energy XUV photons to access photoexcited states at the BZ vertices (TR-XUV-ARPES) [32-35], and high spatial resolution to study the typical micron scale TMD samples (XUV-µ-ARPES) [8]. Here, we overcome these experimental challenges in a single platform to perform TR-XUV-µ-ARPES providing the first direct visualization of dark excitons in a $WSe_2$ monolayer. We report on dark exciton formation pathways under different photoexcitation conditions, the nature of their spectral degeneracy relative to bright excitons, and the dominant role they play in the quasi-equilibrium distribution at long time delays. Our experiments represent a milestone in studies of photo-excited states by providing a global view over the entire Brillouin zone along with unique insight inaccessible otherwise.

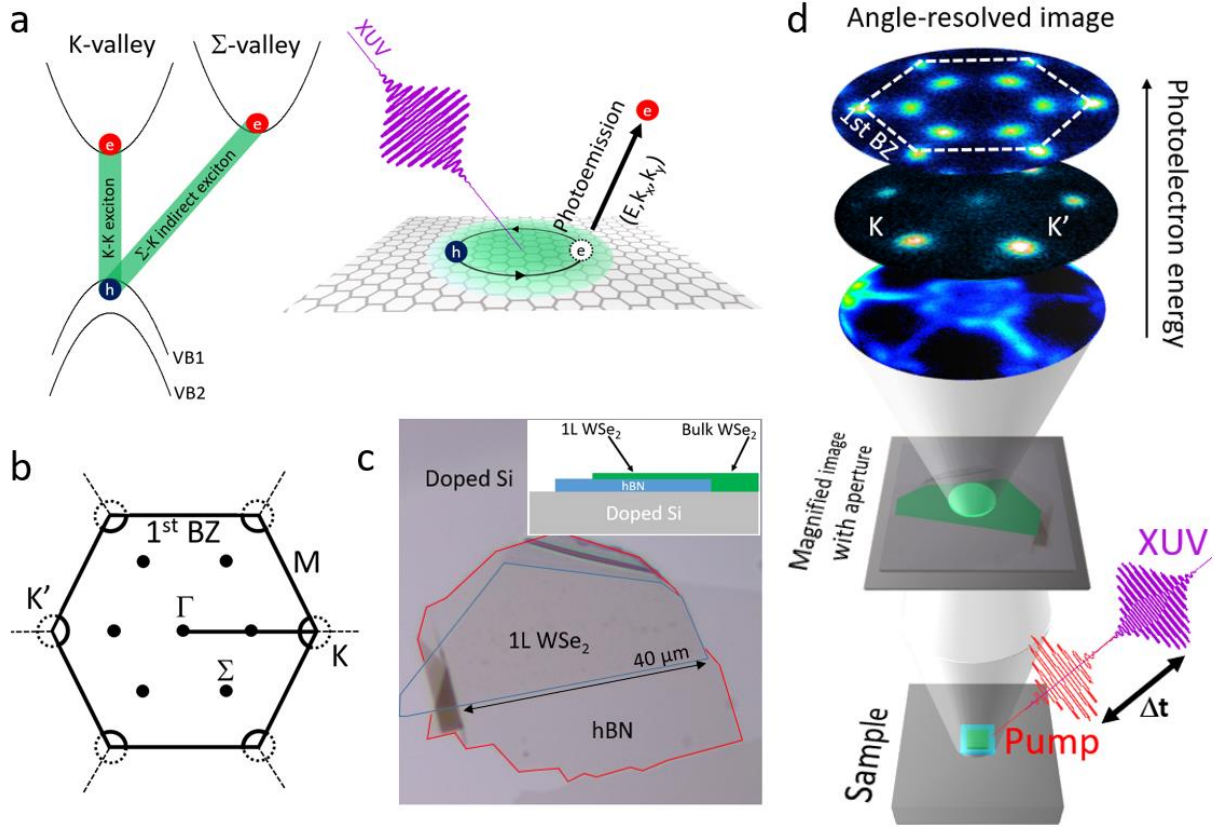

**Fig. 1: Time-resolved-XUV-µ-ARPES of excitons in WSe$_2$ monolayer**. (a) Left: Diagram showing the configuration for K-K direct excitons and Σ–K indirect excitons with holes located in the K-valleys and electrons in K- and Σ-valleys, respectively. Right: Representation of exciton photoemission process. XUV photons photoemit electrons leaving holes from the bound electron-hole pairs. (b) Schematic of the k-space structure of monolayer WSe$_2$ showing the first BZ composed of 6 Σ-valleys and 2 K-valleys (each K-valley is shared between 3 BZs, thus only a 1/3 of each falls within the first BZ) (c) Optical image of the sample composed of a monolayer WSe$_2$ (blue outline) on hBN (red outline) on an *n*-doped Si substrate. Inset: Side view of the sample. (d) Schematic of the experimental setup.

We studied an exfoliated WSe$_2$ monolayer placed on an hBN buffer layer supported by a Si substrate (Fig 1c). The sample was probed at a temperature of 90 K under ultrahigh vacuum conditions. (Details about sample preparation and characterization are provided in Methods and SI §1.) Our experiments are enabled by a custom-built platform that combines an ultrafast, table-top XUV source with a spatially resolving photoemission electron microscope (PEEM) (see Fig. 1d and SI§2). First, we performed an XUV-µ-ARPES measurement to obtain the bandstructure of the unphotoexcited WSe$_2$ monolayer (Fig. 2a). For this, we used ultrafast XUV probe pulse

(21.7 eV) to photoemit electrons from the sample. Using the high-resolution spatial imaging capabilities of our PEEM apparatus, we isolated photoelectrons emitted from only the monolayer region of the sample (see SI §3). These photoelectrons were then dispersed in energy and imaged in the back focal plane in a time-of-flight detector, thereby resolving the angle (i.e., momentum) and energy (with an energy resolution of 0.03 eV) of each photoelectron [36,37] (See Fig. 1c and SI§4). The measured bandstructure showed the spin-split valence band extrema at the K, K' valleys, and had excellent agreement with theoretical DFT calculations (see Fig. 2a and details in Methods).

Next, to measure the excitonic states of the TMD monolayer, we excited the sample with an ultrafast pump pulse, tunable over the visible and near-infrared range of the spectrum. Then, the ultrafast XUV probe pulse discussed above was introduced at a variable time-delay in order to measure the time-, angle- and energy-resolved photoelectron spectrum from the WSe2 monolayer (TR-XUV-µ-ARPES). Recent theoretical studies have predicted photoemission signals from excitons, exhibiting an energy-momentum distribution centered in the corresponding conduction band valley, but binding energy below the conduction band minimum. In our measurements, a striking photoemission signal at positive time-delay after photoexcitation was seen below the bandgap centered at the K (K') and Σ valleys (Fig. 2b). We attribute this signal to the K-K and Σ-K excitons, respectively [19]. To ensure that the detected signals correspond to excitons, we measured the photoemission excitation spectrum (Fig.2c-top panel)), i.e. the integrated the photoemission intensity (from 1 to 3 eV above VB1 in the first BZ) versus the optical excitation energy (varied from 1.58 to 2.85 eV). In the photoemission excitation spectrum, we clearly observed the distinctive A-, B- and C-exciton resonances as previously reported in optical absorption [38]. This energy dependence confirms the dominance of excitons under the experimental conditions used in this work, namely photoexcitation density, sample structure and photoemission probe. Then, we tuned our pump pulse to match the A-exciton resonance as shown in the bottom-panel of fig. 2c. This choice of pump wavelength ensures that we are predominantly and resonantly exciting excitons. Lastly, looking at the energy- and momentum-resolved photoemission signal under these pump conditions, we clearly see a signal in the K-valley, and also at exactly the energy of the A-exciton. This confirms that the photoemission signal at ~1.73 eV, located at ±1.26 Å$^{-1}$ (i.e. in the K-valley) corresponds to the K-K exciton, in agreement with previous theoretical calculations [39] and optical experiments

[40]. By extension, given the expected near-degeneracy of the Σ-K exciton, we attribute the signal at 1.73 eV and ±0.75 Å$^{-1}$ momentum (in the Σ-valley) corresponds to the Σ-K exciton. We also observe the K-K and Σ-K excitons with electrons at the K,K'- and Σ-valleys and presence of holes that can be seen via the depletion of electrons in the K, K' valley by taking the difference between the bandstructures without and after photoexcitation (see supplementary Fig.S12a). The photoemission spectrum taken at different time delays (see supplementary Fig.S12b) after photoexcitation then allowed us to follow the formation dynamics of these dark excitons and learn other aspects of their nature. We note here that in order to eliminate rigid energy shifts or offsets of the entire bandstructure due to surface photovoltage effect or other similar phenomena, we set the peak of the upper spin-split valence band as the zero-energy reference for every time delay (see SI §5 for more details). We also use an optical excitation spot much larger than the sample to eliminate any lateral contribution to surface photovoltage due to local variations of intensity [41].

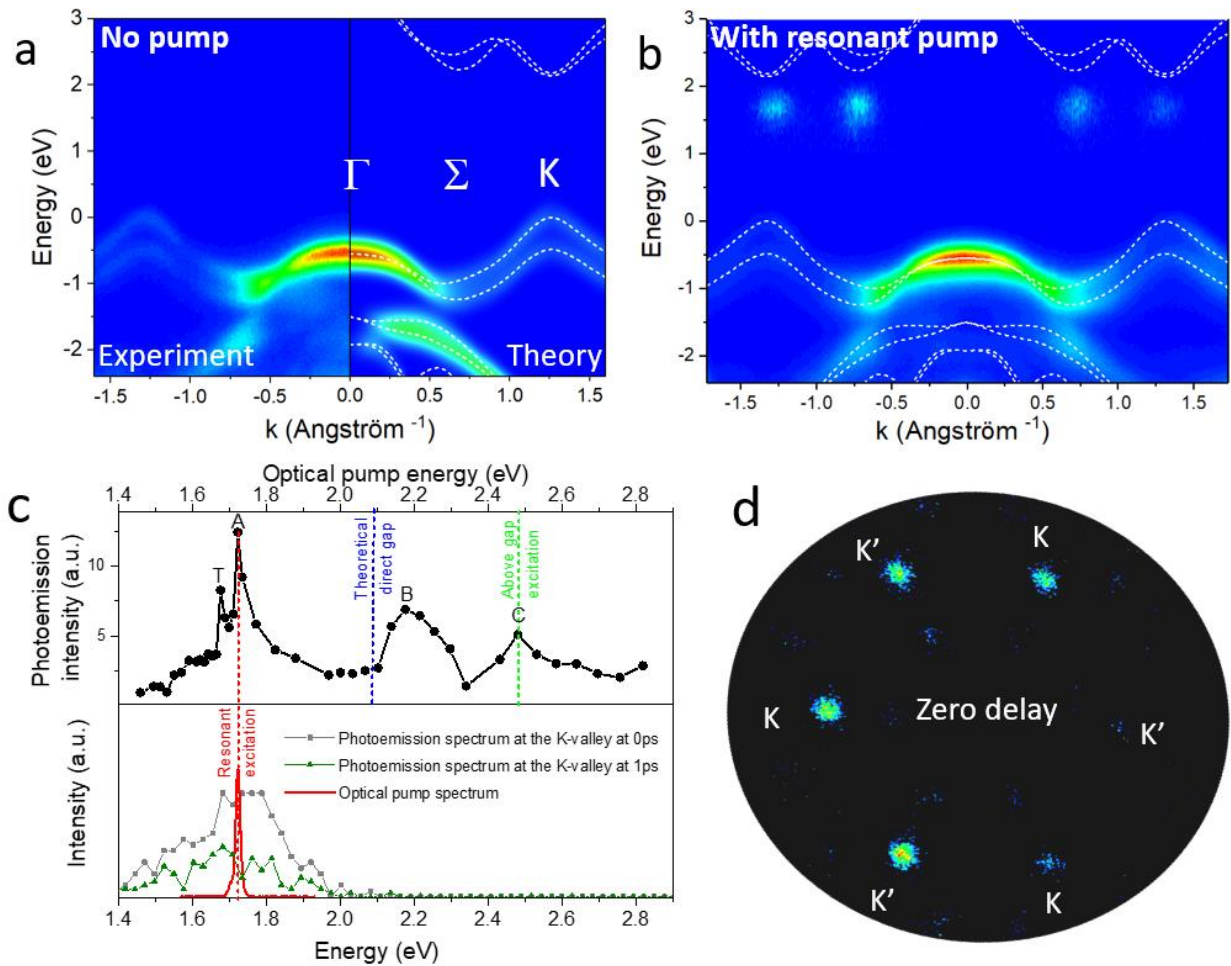

**Fig.2: Excitonic resonances**. Experimental and theory XUV-µ-ARPES results from the WSe2 monolayer without optical excitation. The dispersion of the occupied quasiparticle bands (false color scale of electron emission intensity) is shown together with calculated band structure (dashed white lines). (b) Experimental bandstructure with a resonant photoexcitation at a 0.5 ps delay showing below conduction band direct and indirect excitonic states. (c) Top panel: Photoemission intensity integrated from 1 to 3 eV above the VBM vs optical pump energy. We clearly see the spectrum dominated by resonance at 1.73, 2.17, 2.48 eV, corresponding to the A-, B- and C-excitons of literature. Bottom panel: Spectrum of the resonant optical pump (red) and photoemission energy spectrum at the center of the K-valley at zero-time delay (gray) and also at later time-delays (green). (d) Corresponding momentum-resolved photoemission intensity integrated from 1 to 3 eV above the VBM showing the exciton signals for a resonant excitation in the K-valley at zero-time delay.

To study the formation dynamics, we next resonantly excited the optically allowed K-K exciton as seen in Fig. 2c. We employed linearly polarized pump pulses at 1.72 eV with a fluence chosen to produce an estimated exciton density of $8 \times 10^{12}/cm^2$. At zero-time delay, we observed only the K-K excitons (Fig 3a and 3b). As expected for resonant excitation, the formation of K-K excitons is rapid, as seen by the coincident rise in the photoemission signal and the pump pulse (fig. 3c). The energy of the K-K exciton (1.73 ± 0.03 eV) is consistent with the energy of the pump excitation, within experimental uncertainty, and does not change for longer delay times (fig. 3c). More strikingly, at later times, we see a clear buildup of the dark Σ-K exciton population, at energies nearly degenerate (within our 0.03 eV energy resolution) with the K-K excitons (Fig. 3d). Through the momentum sensitivity of the measurement, we directly observe the formation of the dark Σ-K excitons via scattering from the K-K excitons on a ~ 400 fs timescale (see SI §9). Theoretical studies and indirect optical measurements have reported on phonon-assisted intervalley scattering of excitons [21,40-42], fundamentally different from the electron-phonon interactions seen in bulk TMDs [27]. Our measurements directly access this exciton-phonon scattering and are consistent with the reported timescales of few hundred femtoseconds. The Σ and K signals also show similar recovery time which is consistent with previous studies that the dark Σ-K exciton acts as long-lived reservoir for the K-K exciton [45]. Also of importance is the evolution to a quasi-equilibrium distribution of excitons: We initially create a larger K-K population by resonant excitation, but within a picosecond, the Σ-K exciton population dominates, with K-K/Σ-K ratio tending towards ~0.5 (inset of fig. 3c). We note that in calculating the ratio, we include the population in the entire first BZ comprising 6 Σ-valleys and 2 K-valleys as well as a normalization factor arising from the different photoemission matrix elements (between band states and photoelectron states) at K- and Σ-valleys (see Methods: First Principles Calculations for details). Assuming this limiting ratio reflects equilibrium at the lattice temperature of 90K (and assuming a density of state factor on the order of 1 – details provided in SI §7), one can obtain a tighter bound of <0.015 eV on energy difference of the two exciton species. A video of the exciton dynamics measured over the full BZ after resonant photoexcitation is presented in the SI.

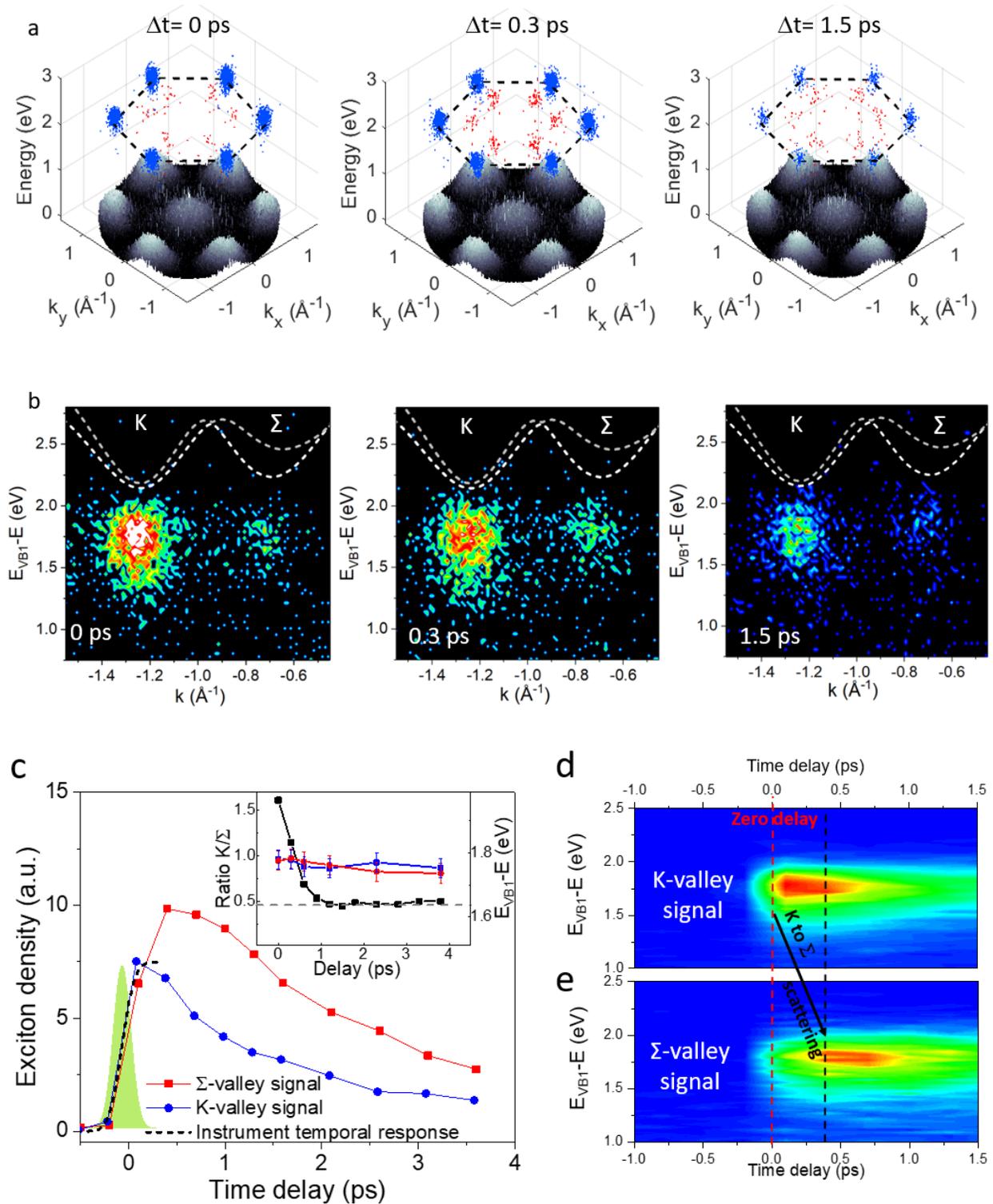

**Fig. 3: Exciton dynamics following resonant excitation.** (a) TR-ARPES data for delay times of 0, 0.3, and 1.5 ps for the full 2D BZ (in plane) and emission energy (vertical). The top of the valence bands is displayed in grayscale, while emission from the K-K (Σ-K) excitons is plotted in blue (red) dots (see SI §4). The black dashed line defines the boundary of the 1st BZ. (b) TR-ARPES data for delay times of 0, 0.3, and 1.5 ps along the line connecting the K

and Σ valleys. The dotted lines show the calculated (spin-split) conduction bands. (c) Exciton density versus time delay at Σ-valleys (red) and K-valleys (blue). The exciton density was determined by the ARPES signal integrated over the 2 K-valleys or 6 Σ-valleys of the 1st BZ and an energy range from 1 to 2.5 eV, with a correction factor for the respective photoemission matrix elements. The dotted black line shows the instrument response function, i.e. the convolution of the pump and probe pulse. In green, we plot the corresponding gaussian pulse. Inset: Ratio of the K/Σ population from Fig. 2c shows dominant K-K exciton population initially, but a ratio approaching ~0.5 at longer time-delays. (d) Time-resolved spectrum of the K-valley signal integrated over the first BZ. (e) Time-resolved spectrum of the Σ-valley signal integrated over the first BZ. K-K excitons are populated directly upon photoexcitation. We observe a clear delay in the rise of the Σ–K population due to the K to Σ scattering.

Finally, we turned our attention to the dynamics after above bandgap excitation. We used a 2.48 eV linearly polarized pump pulse to excite carriers well above bandgap, producing an estimated carrier density of $2 \times 10^{12}/cm^2$. Surprisingly, the exciton relaxation pathways, particularly for the dark Σ-K excitons, were dramatically different from those observed for resonant excitation. Fig. 4a shows snapshots of the full ARPES data at different time delays at the K and Σ valleys (the full movie of the exciton dynamics is included in SI). Immediately after excitation (zero delay), we see a broad distribution in the K- and Σ-valley centered at an energy of ~1.90 eV (see Fig. S10 of SI) that could involve contributions from both free carriers and excited excitonic states. Future experiments with improved time and energy resolution are needed to explore these very early dynamics (see SI for a more detailed discussion). Beyond this regime, we observed full relaxation into the K-K or Σ-K excitons within 500 fs (Fig. 4b). Previously, optical and mid-infrared spectroscopic measurements had reported the sub-picosecond formation of exciton dynamics [46,47], but lacked separate access to the dynamics of the different types of excitonic states, such as the dark Σ-K excitons. The relaxation process can also be described by plotting the average energy of the photoemission signal versus time (fig. 4c), giving an energy relaxation time of 500 fs. Beyond 500 fs, the peak energy of the distribution at the Σ- and K-valley remains constant at approximately 1.73 eV, matching the exciton energies under resonant conditions (Fig 3c Inset and 4c). A striking departure from the resonant excitation case is that Σ−K excitons appear coincident with the K-K excitons (rather than at a finite delay after scattering of the K-K excitons as for the resonant excitation). We also observe that the dark Σ−K exciton density dominates the bright K-K density at all time-delays (Fig. 4c and 4c inset), in contrast to the resonant excitation case. However, at long time delays, under both resonant and above-gap excitation, the system evolves to a similar quasi-steady state, with a nearly identical K-K/Σ-K exciton population ratio, and nearly degenerate exciton energies with respect with to the valence band maximum. The exciton binding energy is given by the difference between the conduction band minimum and the energy of the constituent electron photoemitted from the center of the valley, with a correction for the exciton thermal energy [17]. At these long time-delays, where quasi-equilibrium is reached, we can assume exciton temperature to be close to the lattice temperature (90 K) [44] leading to an estimated exciton thermal energy of ~7 meV. With this thermal energy being much smaller than our expt. resolution (see Methods section), we estimate binding energies without explicitly accounting for the thermal correction as ~390 meV and ~480 meV (with respect to the conduction band minima) for the K-K and Σ-K excitons, respectively. While the former can be compared to results of various optical spectroscopy measurements [3], the binding energy and momentum-space distributions of the dark exciton in monolayer TMDs are not easily accessible to other experiments [21].

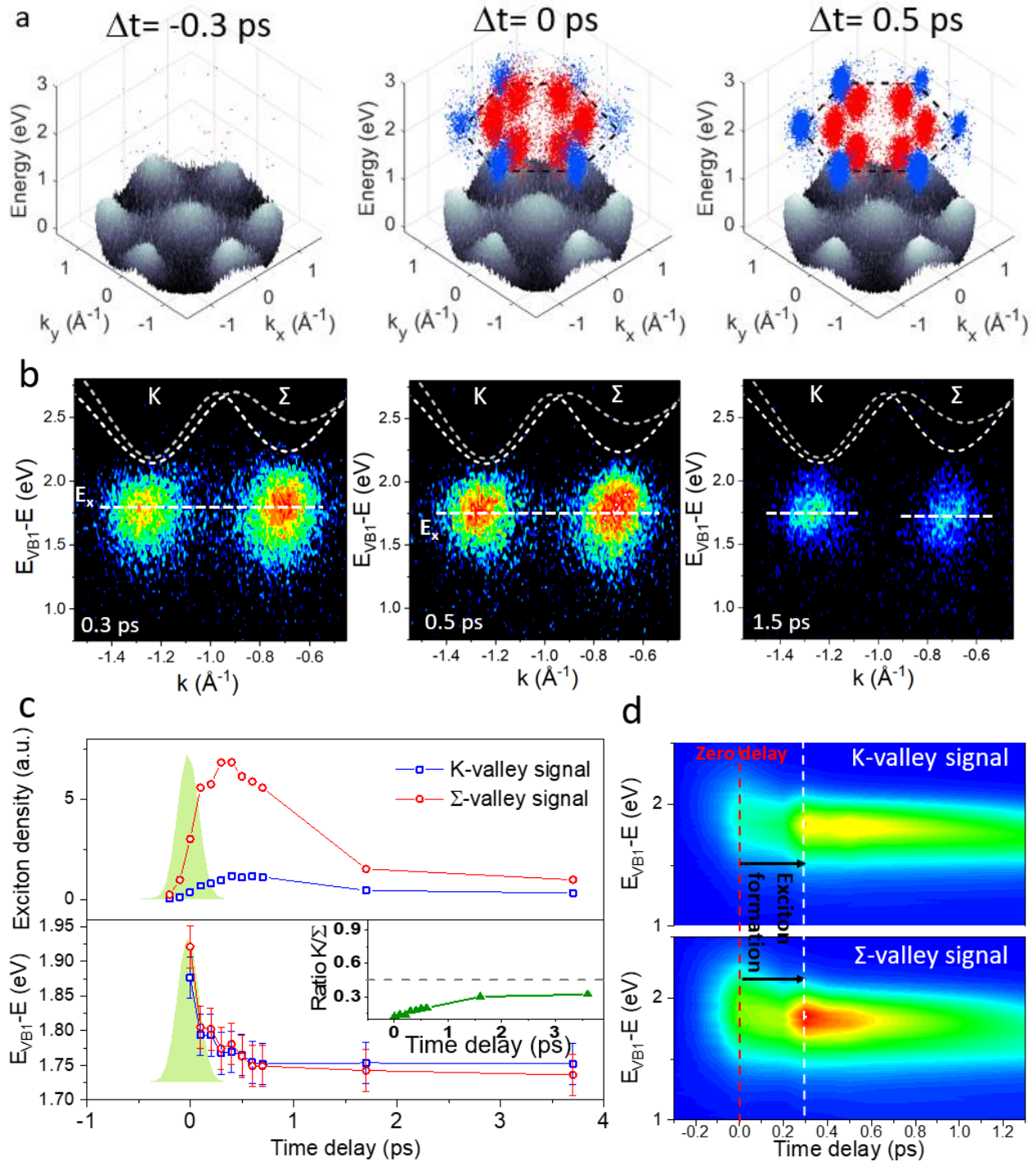

**Fig. 4: Exciton formation and dynamics following above-gap excitation.** (a) 3D plots of the experimental data at 0 ps, 0.3 ps and 1.5 ps time delays. The top of the valence bands is displayed in grayscale. The K-K (Σ-K) excitons are plotted in blue (red) dots. The black dashed line defines the boundary of the 1st BZ. (b) ARPES data at the K- and Σ-valley for 0.3 ps, 0.5 ps and 1.5 ps time delays. In dotted line is theoretical calculation of the conduction bands. (c) upper panel: K-valleys (blue) and Σ-valleys (red) exciton density over the first Brillouin zone from 1 eV to 2 eV. Lower panel: Center energy of the photoexcited population obtained from fits with Gaussian function at the center of the K- and Σ-valley. Inset: Ratio of the total populations between K- and Σ-valleys extracted from upper panel. (d) Time-resolved spectra at K- (upper panel) and Σ-valleys (lower panel) showing the formation and relaxation of excitons. (e) Photoemission energy spectra at the K valley from -0.3 ps to 3.6 ps time delays showing the evolution of the valence bands and photoexcited populations.

Our measurements, using TR-XUV-µ-ARPES to access strongly-bound, few-particle excitonic states in 2D semiconductors and their dynamics, open new possibilities. Such direct access to dark excitons, or other valley- and spin-polarized excitons will enable their utility in quantum information [48], valleytronic and spintronic schemes and in creating novel many-body excitonic states [9, 10]. Energy- and momentum-resolved photoemission studies of excitons could directly image excitonic wavefunctions in real- and momentum-space. Via the energy-momentum dispersion relationship, one could measure important physical properties, such as the kinetic energy and temperature of photoexcited excitons [18-21]. Future measurements could access few-particle excitations, such as trions, biexcitons and intervalley-excitons in TMD heterostructures, which may be expected to have their own unique photoemission signatures. Lastly, we expect our measurements to extend to other condensed matter systems in providing a complete picture of the transformation of their electronic structure along energy & momentum axes after optical illumination.

**Methods**

**Sample preparation.** The $WSe_2$ monolayer (40x20 µm) and hBN were mechanically exfoliated and transferred onto a *n*-doped Si substrate. The hBN buffer layer was ~20 nm thick. This thickness was chosen to be great enough to prevent rapid quenching of excitons by energy transfer to the substrate, but also small enough to provide conductive channels to the Si substrate under excitation by the XUV probe, and thus prevent charging of the sample. The monolayer was also in contact with bulk $WSe_2$ directly on Si, which could a provide an alternate conductive pathway.

**Optical pump and XUV photoemission probe.** We used a laser system based on a Yb doped fiber amplifier operating at 1 MHz and providing 230 fs pulses at 1030 nm with a pulse energy of 100 µJ. 20 µJ were used to pump an optical parametric amplifier with a 5nm spectral bandwidth, tunable from 320 to 2500 nm with pulse energies in the range of 0.1-1 µJ. Optical pulses from this source served to photoexcite the sample with a *p*-polarized beam at a 22° angle of incidence. The pulse duration at the sample was consistently measured to be ~240 fs for the wavelength range used in this experiment. 70 µJ of 1030 nm were converted into 515 nm radiation with a BBO crystal providing 30 µJ to generate XUV radiation. XUV radiation was produced via high harmonic generation (HHG) by focusing the beam to an intensity of $3\times10^{14}$ W/cm² in a Kr gas jet located in a vacuum chamber in a similar manner as Buss et al. [26]. The photon flux at 21.7 eV (measured with a calibrated XUV CCD camera) was $>10^4$ photons/pulse. This excitation led to an average of 0.1 photoelectron measured at the detector per pulse, i.e., to $10^5$ photoelectrons/s.

**Sample photoexcitation.** For the resonant study, the sample was photoexcited at 1.72 eV (FWHM 5nm) with a fluence of 14 µJ/cm², corresponding to an estimated density of excitons of $8 \times 10^{12}$ cm$^{-2}$. For the off-resonant case, we used 5 µJ/cm² excitation at 2.48 eV (FWHM 5nm) creating an estimated initial photocarrier density of $2 \times 10^{12}$ cm$^{-2}$. In both cases, we assumed a typical absorption of 15% from the monolayer $WSe_2$, as previously reported.

**Time-resolved XUV micro-angle resolved photoemission spectroscopy (TR-XUV-µ-ARPES).** TR-XUV-µ-ARPES was performed in a time-of-flight momentum microscope (see SI §2 for a detailed description). Photoelectrons emitted from the sample are collected by an immersion objective lens giving access to the full half-space above the sample surface. Momentum space maps of selected sample area were obtained by imaging the Fourier plane at the back of the objective lens. By inserting a field aperture at the Gaussian image plane of the microscope, we selected a region of 6x40µm on the monolayer $WSe_2$ from which we collected the photoelectrons. The kinetic energy of the photoemitted electrons is measured by a time-of-flight detector, the effective energy and momentum resolution of our system is set to be around 30 meV and ~0.01 Å$^{-1}$, respectively (see SI §4).

**First-principles calculations**. Density functional calculations (DFT) within the local density approximation (LDA) were performed using the Quantum ESPRESSO package [37]. We used the experimental lattice constant of 3.28 Å in our calculations. The GW [38] calculations were carried out using the BerkeleyGW package [39]. In the calculation of the electron self-energy, the dielectric matrix was constructed with a cutoff energy of 35 Ry. The dielectric matrix and the self-energy were calculated on an $18 \times 18 \times 1$ k-grid. The quasiparticle bandgap was converged to within 0.05 eV. The spin–orbit coupling was included perturbatively within the LDA formalism. The calculations of the ARPES efficiencies for different regions of the BZ were performed by using the quasiparticle band structure, with the intensity given by the modulus squared photoemission matrix elements. The matrix elements between the Kohn-Sham wavefunctions (initial states) and the free electrons wavefunctions (final states) were calculated within the electric-dipole approximation. The broadening was set to 0.2 eV with a Gaussian lineshape. In general, the GW method is expected to provide a systematic error of 100 meV for quasiparticle band gaps of semiconductors [52]. Another important consideration is the convergence parameters coming from the number of k points, number of empty bands, and dielectric cut-off which in our case yield an error smaller than 50 meV.

**Acknowledgments**

This work was supported by JSPS KAKENHI Grant number JP17K04995 and in part by funding from the Femtosecond Spectroscopy Unit, Okinawa Institute of Science and Technology Graduate University. We thank OIST engineering support section for their support. This research (theoretical analysis and first-principle calculations) was partially supported by NSF through the University of Washington Materials Research Science and Engineering Center DMR-1719797. Analysis at SLAC was supported by the AMOS program, Chemical Sciences, Geosciences, and Biosciences Division, Basic Energy Sciences, US Department of Energy under Contract DE-AC02-76-SF00515. T.C. acknowledges support from the Micron Foundation and a GLAM postdoctoral fellowship at Stanford. Computational resources were provided by Hyak at UW, and the Extreme Science and Engineering Discovery Environment (XSEDE), which is supported by National Science Foundation under Grant No. ACI-1053575. The work of M.C. and X. L. at Austin was partially supported by the National Science Foundation through the Center for Dynamics and Control of Materials: an NSF MRSEC under Cooperative Agreement No. DMR-1720595.

## Contributions

J. M., M.K.L.M. and K.M.D. designed the experimental setup. J.M., M.K.L.M. and C.S. built it. J.M., M.K.L.M., C.S., A.A.M. and E.L.W. performed the experiments. J.M., M.K.L.M., A.A.M. and N.S.C. analyzed the data. J.M., B.M.K.M, C.S. and E.L.W. designed and built the XUV source. V.P. and A.K. characterized the sample. M.C. and X. L. prepared the sample. T.C. and T.F.H. provided theoretical support. T.C. performed first principle theoretical calculations. K.M.D. supervised the project. All authors contributed to discussions and manuscript preparation.

## Competing interests

J.M., M.K.L.M. and K.M.D. are inventors on a provisional patent application related to this work filed by the Okinawa Institute of Science and Technology School Corporation (US62/834,829 filed on 26 April 2019). The authors declare no other competing interests.

## Data availability

The data that supports the finding of this work are available upon request to the corresponding author.

# Supplementary Information

1. **Sample geometry and characterization**

Fig.S1a shows an optical image of the sample composed of a 1L-WSe$_2$/hBN heterostructure on a n-doped Si substrate. The heterostructure was prepared by dry transfer method using PDMS stamps [S1]. We exfoliate hBN onto a commercially PDMS from Gelpak and identify few layer hBN for transfer via optical contrast. Once we have identified the few layer hBN flake, we transfer it onto the n-doped Si substrate using a home-buit transfer setup. Similarly, we exfoliate WSe$_2$ on a PDMS stamp and identify 1L WSe$_2$ sample using optical contrast. We then transfer the 1L WSe$_2$ onto the previously transferred hBN to make the heterostructure. We then heat the sample in vacuum during 4 hours at 450°C to remove interfacial trapped air. To ensure that the sample did not exhibit any charging during the photoemission process, a bulk region of the sample attached to the monolayer was directly in contact with the conducting Si substrate to provide an extraction channel (Fig. S1b). (We note that the Si substrate has a few nm native oxide layer which does not impact our observations.) Finally, we confirm the successful assembly of the heterostructure sample by measuring the PL from the A exciton of the transferred 1L WSe$_2$ on hBN using a 532 nm CW excitation at room temperature (Fig. S1c). The main A exciton line was measured at 1.64 eV as expected from a WSe$_2$ monolayer at room temperature [S1, S2].

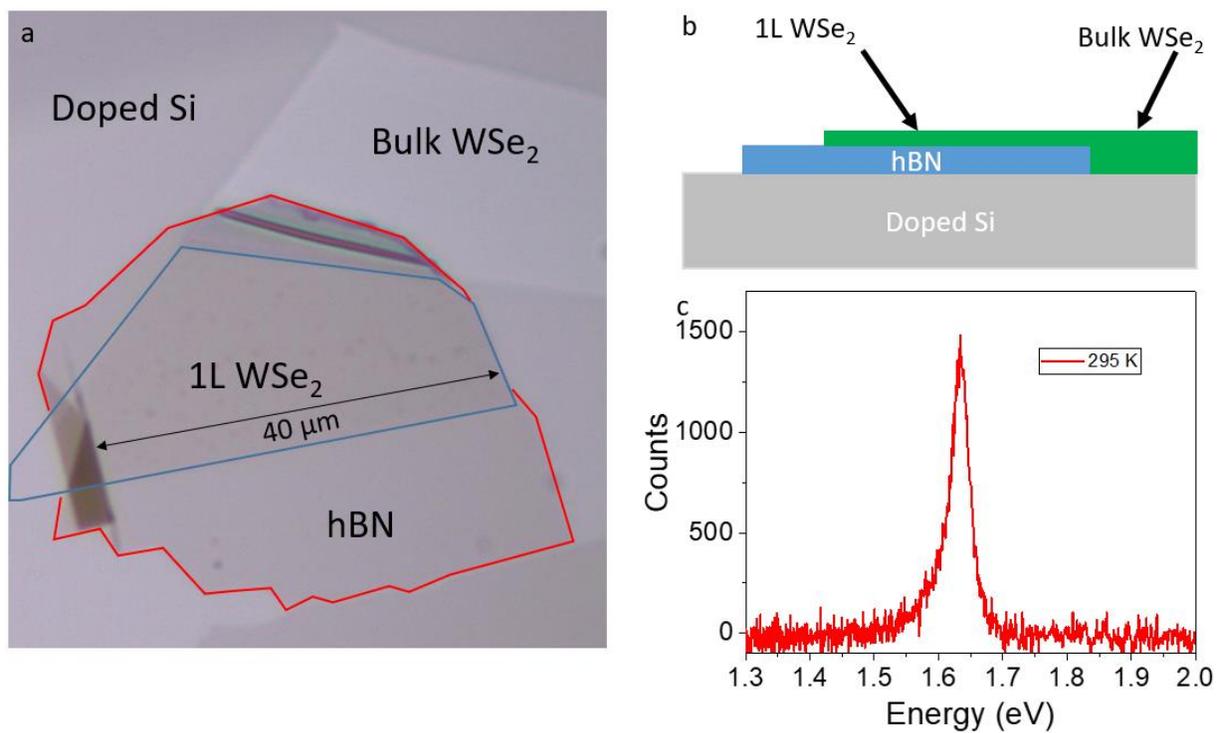

**Fig.S1:** (a) Optical photo of the sample (b) schematic of the sample. (c) Photoluminescence spectrum at 295K of the monolayer WSe$_2$

## 2. Experimental setup

We provide in Fig.S2 a schematic of the experiment.

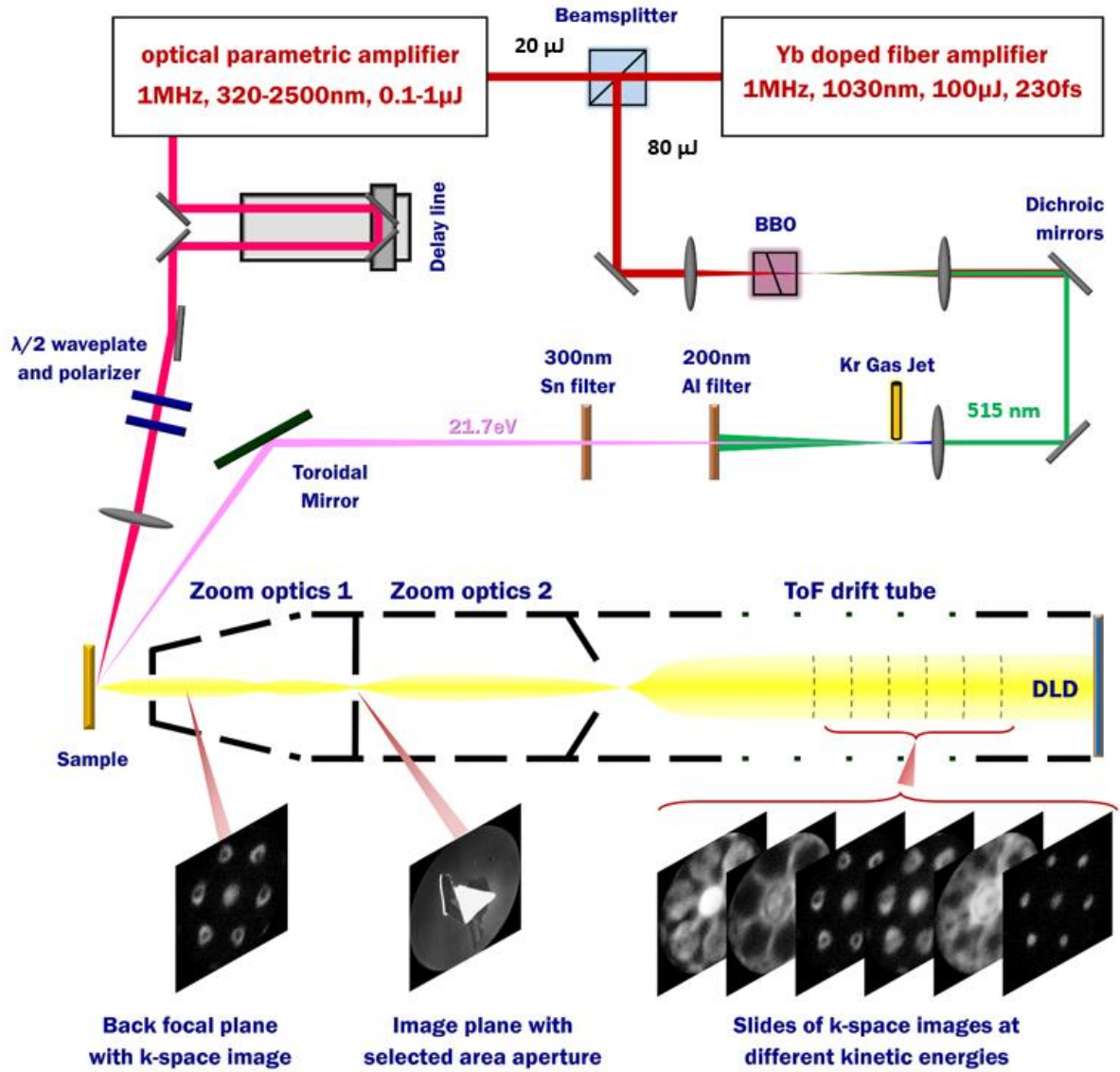

**Fig. S2:** Schematic of the time resolved micro-angle resolved photoemission spectroscopy (tr-µ-ARPES) setup.

(a) <u>Optical Pump</u>

We use an Yb fiber amplifier operated at 1MHz and delivering 100 µJ pulse energy at 1030 nm. 20 µJ are used to pump an optical parametric amplifier operated at 1 MHz, providing narrow linewidth (5 nm or 12 meV FWHM at 720 nm) and tunable from 320 to 2500nm. A delay line is used to change the time delay between the optical pump and the XUV probe. A half waveplate is then used to control the polarization of the pump that is then focused into the microscope chamber via one the port with a fused silica window.

(b) <u>XUV probe</u>

80 µJ from the source laser is frequency doubled by a 500µm-thick BBO crystal providing ~40% efficiency. 35 µJ of converted 515 nm beam is then focused into a vacuum chamber on a Kr gas jet (1 bar backing pressure) with a power density of $3 \times 10^{14}$ W/cm². High harmonic lines are generated and then filtered with a 200 nm-thick unsupported Al foil filter acting as a shortpass filter transmitting photons with energies higher than 15 eV. A 300 nm-thick Sn filter is then used as a bandpass filter to select the 21.7 eV harmonic. The harmonic spectrum and photon flux were characterized in-situ by positioning a SiC mirror to divert the XUV on a characterization setup comprising a diffraction grating (500 grooves/mm) and a XUV CCD camera. Fig.S3 shows the measured spectrum consisting in a main harmonic at 21.7 eV. The Sn bandpass filter calculated transmission (CXRO x-ray database) is displayed showing that only the 21.7 eV harmonic is selected. Finally, we use a f=500 mm toroidal mirror to collect and refocus the XUV at a 4° grazing angle in a 2f-2f configuration on the sample (see section 3).

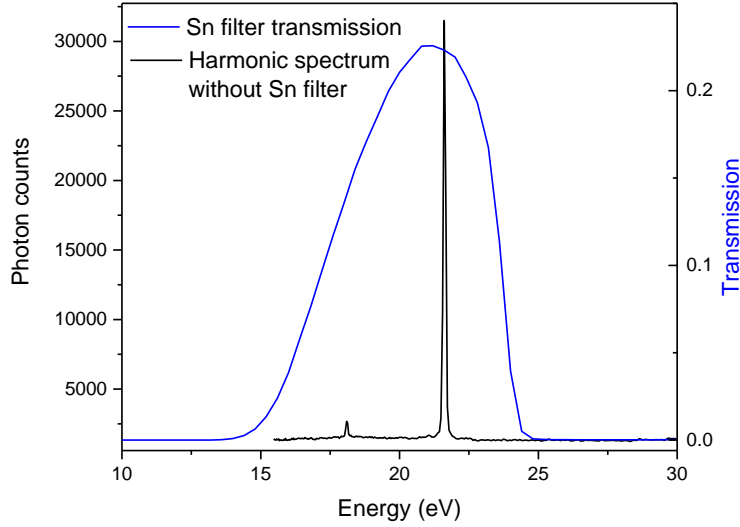

**Fig.S3:** Generated harmonic spectrum (black) and Sn filter transmission (blue).

(c) TR-XUV-μ-ARPES

TR-XUV-μ-ARPES was performed in a time-of-flight momentum microscope (Metis 1000, SPECS GmbH), A simplified schematic of our setup can be found in the SI §2 Fig.S2, for other details about the microscope please refer to the following references [S3–S5]. In the momentum microscope, photoelectrons emitted from the sample are collected by an immersion objective lens and it gives access to full half-space above the sample surface. Unlike normal ARPES apparatus, in which high spatial resolution requires tightly focusing of the photon beam, in the momentum microscope, photoemitted electrons from a small area (down to few μm) of the sample can be selected by placing a field aperture at the image plane of the first zoom optics. The second zoom optics can operate in two different ways, 1) in imaging mode: further magnify the real space image, and 2) in ARPES mode: project the momentum space image at the entrance of the time-of-flight (TOF) drift tube. In imaging mode, the microscope provides spatial resolution better than 50nm. It facilitates identification of region of interest, and accurate placement of the field limiting aperture. In the field free region of the drift tube, electrons of different kinetic energy are separated out and positions of these electrons are recording by a 2D delayline detector (DLD). This setup allows collection of 3D ARPES data, simultaneously recording the (kinetic energy, $k_x$, $k_y$) of the photoemitted electrons.

3. **Spatial selection of the monolayer area for photoemission**

In Fig. S4, we show the result of spatial selection for photoemission. Fig.S4a shows an image of the sample obtained with a broadband Hg lamp with a large excitation and no aperture. By using a combination of a hard physical aperture in the image plane and a focused XUV spot, only the monolayer area of the sample contributes to the photoemission as shown in Fig.S4b with the imaging of the sample with the XUV probe.

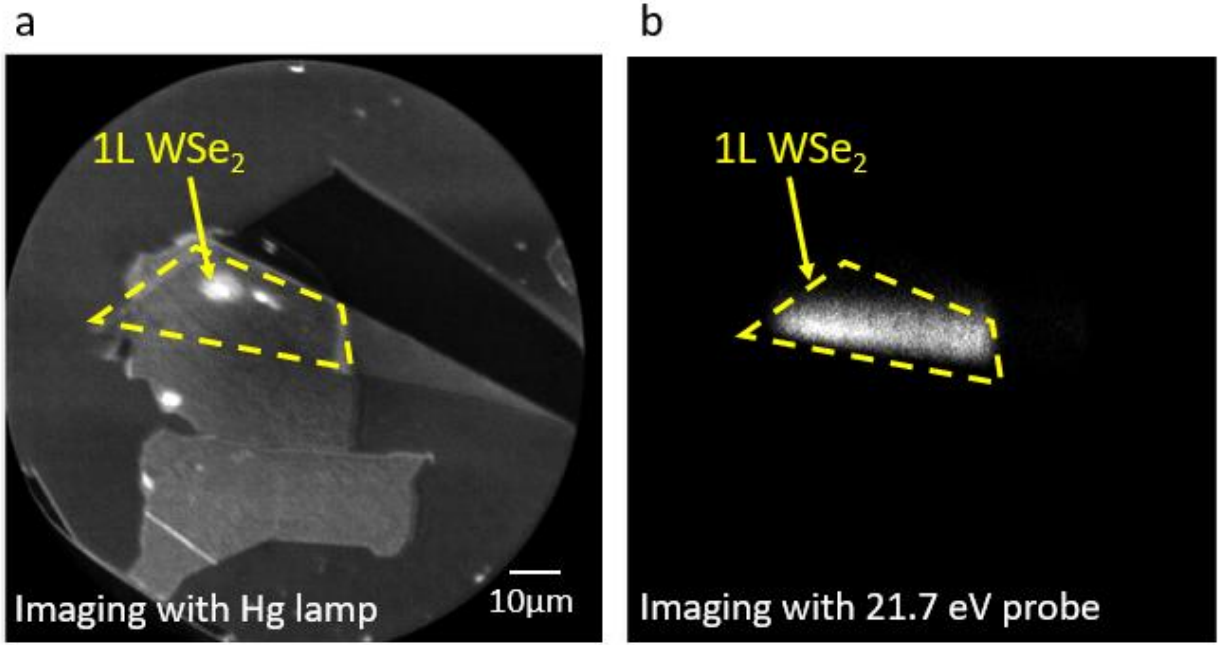

**Fig.S4:** (a) Image of the sample using a Hg lamp. The monolayer area is shown in dashed yellow lines.(b) Image of the sample using the 21.7 eV probe with an aperture showing that only the monolayer area contributes to the photoemission.

4. **Instrument energy and temporal resolution**

The energy resolution of the instrument depends strongly on the drift energy in the drift tube. In our experiment, we calibrated the energy resolution of the instrument by measuring the Fermi edge (16-84%) of clean Au(111) surface at 90K, we obtain an effective energy resolution of 30meV (Fig.S5-left). To ensure that photoelectrons and our energy spectrum are not affected by Coulomb interaction or vacuum space charge effects [S6–S8], we checked the photoelectron spectrum of Au(111) under different photon flux, we collect from $1 \times 10^4$ to $2 \times 10^5$ photoelectrons per second and we do not see any detectable spectral broadening or spectrum shift (Fig.S5-right). It confirms that, with electron counts lower than the laser repetition rate (1MHz), we are generating less than 1 photoelectron per pulse, and hence we are working below the space charge limit. Similarly, sample charging effects would have resulted in additional spectral shift and broadening in Fig. S5 (right) with increasing probe intensity (i.e. detected photoelectron count rate) which is not observed here.

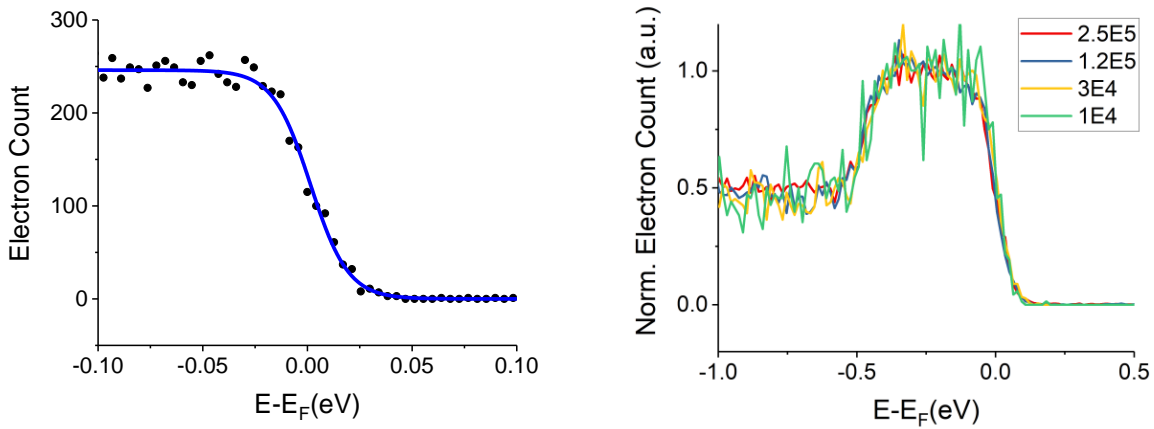

**Fig.S5:** (left) Energy spectrum at the valence band maximum of the K-valley with no optical pump and at negative delay for a 2.48 eV pump. (right) Investigation of the space charge effect by measuring the Fermi edge of Au(111) as a function of the number of photoelectrons detected per seconds. No obvious spectral broadening or shift can be detected.

We have measured the instrument temporal response to ensure that the observed dynamics can be resolved. To do so we have performed an autocorrelation of the optical (Fig. S6a) showing a Gaussian pulse duration of 241 fs. We then plot the pulse error function corresponding to the measured autocorrelation and confirm that, in the case of resonant excitation, the rise time of the measured signal in the K-valley is pulse width limited (Fig.S6b).

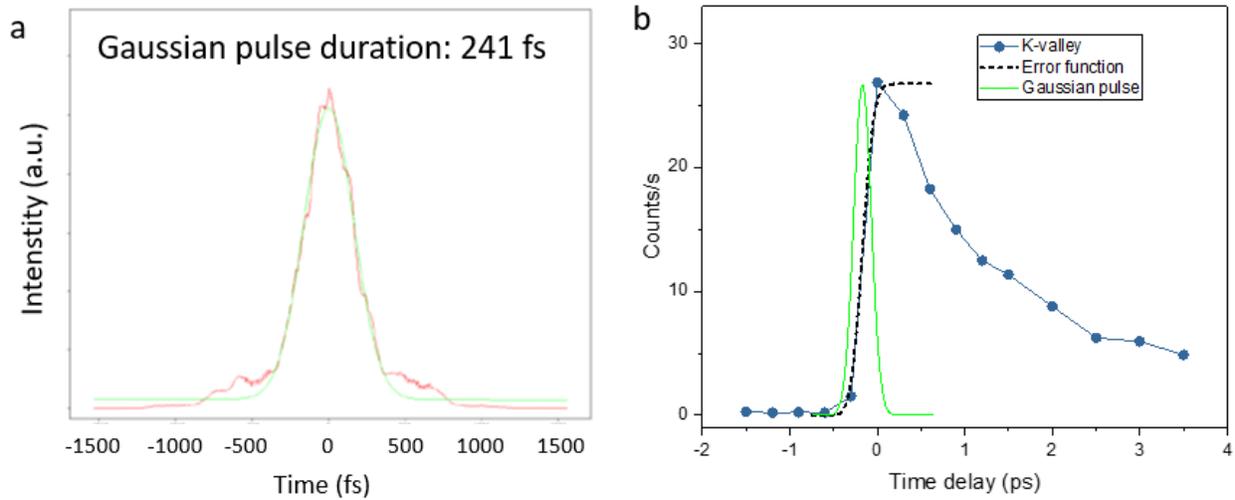

**Fig.S6:** (a) Autocorrelation of the optical pump (red) and Gaussian fit (green). (b) Dynamics of the exciton population in the K-valleys for a resonant excitation (blue), Gaussian pulse duration (green) and corresponding pulse error function (dashed black line).

5. **Energy spectra, bandstructure plots and energy reference**

Fig.S7a displays a ($k_x$,$k_y$) image for the above bandgap excitation case at an energy cut corresponding to the center of the excitonic states (1.73 eV). The K-K' points are indexed from K1 to K6. In order to plot both the energy spectra and the dispersion graphs, we took a cut along a K-Γ-K' axis as depicted the Fig.S7a with a 10 pixels width. The result of the cut is shown in Fig.S7b (K- Γ) with an overlay of the calculated bandstructure. For all data presented in this work, including each time delay, the energy reference was defined by the peak of a Gaussian curve fitting to the top valence band (VB1) at K as shown in Fig.S7c.

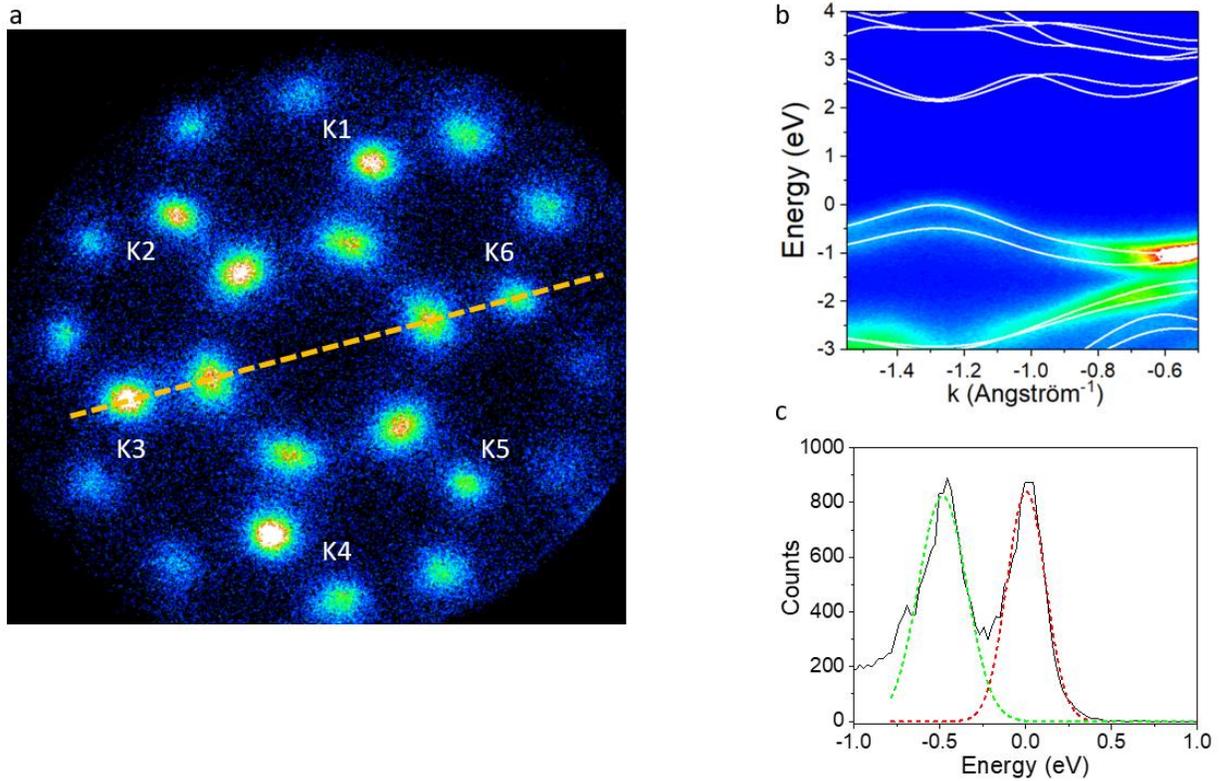

**Fig.S7:** (a) (kx,ky) image at the exciton energy. The K points are labeled from K1 to K6. (b) Extracted cut corresponding to the orange dashed line in (a) with theoretical bandstructure. (c) Energy spectrum obtained with a cut at the top of the K point valence band. The spin-split valence bands are fitted with Gaussian functions to define the zero energy reference.

In Fig.S8a, we show the difference between uncorrected and corrected data for energy vs time delay integrated in the K-valley for an above gap excitation when using the method described above to set the energy reference to the peak of the top valence band in the K-valley. In Fig.S8b, we show the corresponding energy correction for each time delay applied to rigidly shift the bandstructure so the top valence peak is at zero. This energy correction does not exceed 60 meV.

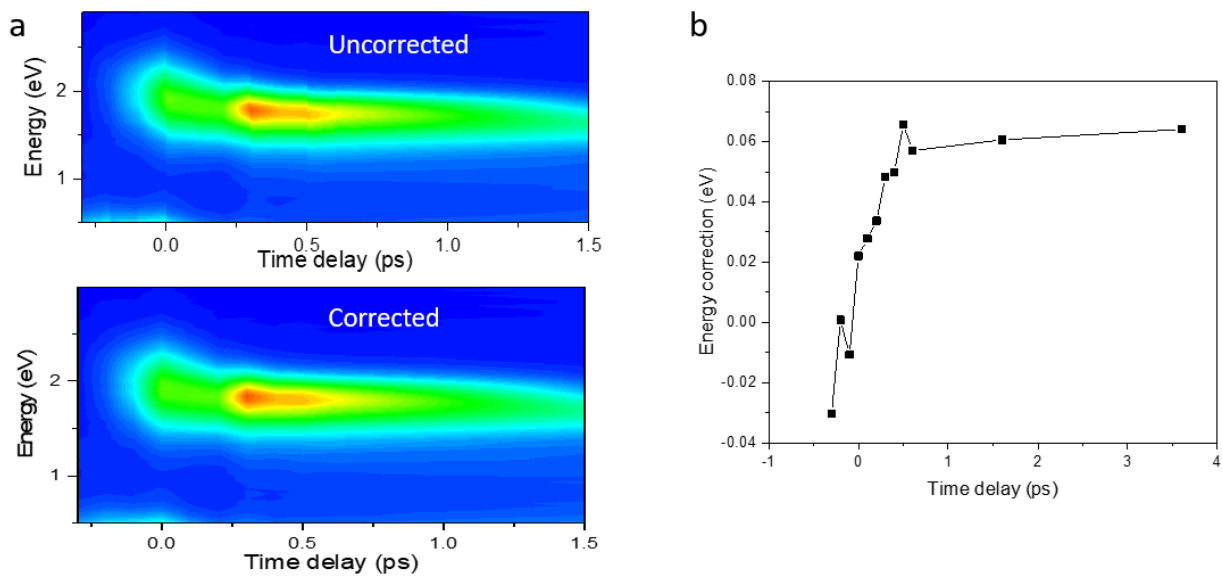

**Fig.S8:** (a) Uncorrected (top) and corrected (bottom) integrated signal vs time delay in the K-valleys for an above gap excitation. (b) Corresponding energy correction to the bandstructure for each time delay.

6. 3D representation of the experimental data

We used a homebuilt Matlab code to plot the 3D figures (3a and 4a) presented in the manuscript. Figure S9a shows the data plotted for the resonant excitation case. This dataset was taken with a 2 minute integration time for each time delay. For clarity, we performed a 60° degrees rotating summation centered at the Γ point. We note that this was performed only for the 3D representation in figure 3a of the resonant excitation case and that the rest of the data presented in the manuscript are presented as-is with no post-treatment. Fig. S9b shows the data plotted for the above-bandgap excitation. Here, we used a 3 hour integration time for each time delay.

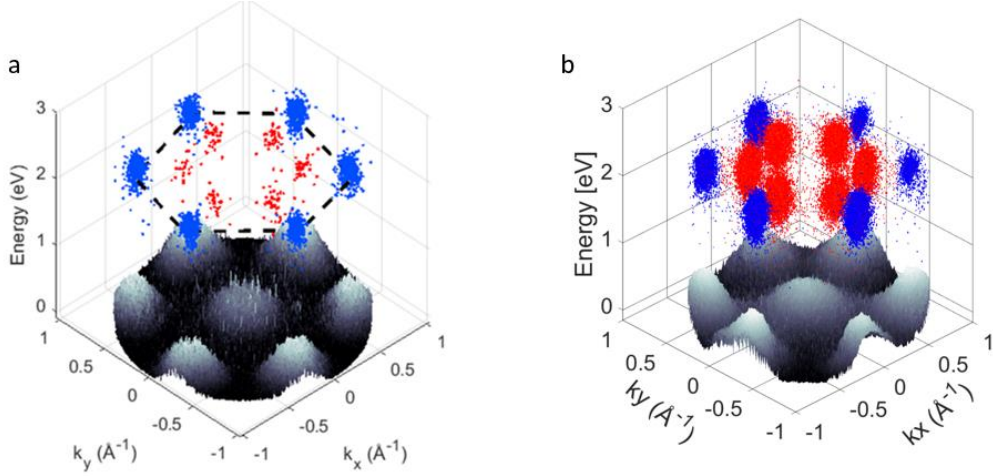

**Fig.S9:** 3D plot for (a) the resonant excitation case with 2 minute integration time and 60° rotating averaging and (b) the above bandgap excitation case with 3 hour integration time.

Using the same method, we generated videos by adding successively the 3D plots for each time delay.

Video 1: Resonant excitation – Time delay from -1.5 ps to 3.5 ps

Video 2: Above-gap excitation – Time delay from -0.3 ps to 3.6 ps

7. **Calculation of the population ratio between K- and Σ-valley excitons**

In this section, we use the Boltzmann statistics to calculate the population ratio between excitons at K and Σ valleys in the quasi-equilibirum conditions after the initial cooling. Under such conditions in the experiment, the exciton population in either the K or the Σ valley is smaller than $10^{12}/cm^2$, which is below the density threshold for forming biexciton, exciton liquid, or the Mott transition. With the Boltzmann statistics, the population ratio between K and Σ valleys is related to three factor, the energy difference between the two excitons $\Delta E$, temperature $T$, and the exciton density of states of the two excitons $\frac{\rho_K}{\rho_\Sigma}$.

At 2D, the exciton density of states ($\rho$) are related to the exciton effective mass and the number of different species. For a Wannier exciton in a hydrogen model, the exciton effective mass is the total mass of the electron and the hole, which is close to $m_0$ (electron rest mass) for both K-K and

K-Σ excitons. Since $\rho_K$ includes contribution from 4 different species (K-K, K'-K', K-K', and K'-K), and $\rho_\Sigma$ includes 12 different species (K-$\Sigma_{1\sim6}$, K'-$\Sigma_{1\sim6}$), $\frac{\rho_K}{\rho_\Sigma} = 3$.

A correction to this ratio can arise from the effective mass enhancement due to the exciton binding. Previous studies have found a ~ 30% increase due to the attractive Coulomb interaction between the quasielectrons and quasihole [S9]. However, such an increase is directly related to the ratio between the exciton binding energy and the quasiparticle bandwidth, which is almost the same at K and Σ valleys [S10]. Therefore, the correction will increase $\rho_K$ and $\rho_\Sigma$ to the same extent, and will not affect their ratio.

As a result, using $\frac{\rho_K}{\rho_\Sigma} e^{-\frac{\Delta E}{k_B T}} = 0.5$, we get $\Delta E = -3$ meV. The uncertainty of this estimation arises from the spin degree of freedom, which may add additional phase space to K-K excitons. However, such correction is small and would not change the energy difference by a factor of 2.

8. **Dispersion analysis for above gap excitation at zero time-delay**

For the above gap excitation, the observed signal at zero time-delay is complex and our experimental temporal resolution does not allow to distinguish among the various possible contribution in the early dynamics. In Fig. S10-left panel, we observe a positive dispersion at zero time delay that could originate from free carriers in the conduction band, hot exciton states and include effects such as bandgap renormalization. In Fig. S10-right panel, we show Gaussian fits of the energy distribution performed at different k-momenta from -1.5 to -1.0 Å$^{-1}$. From that simple analysis, it clear that the overall distribution exhibits a positive dispersion with a minimum around 1.75 eV.

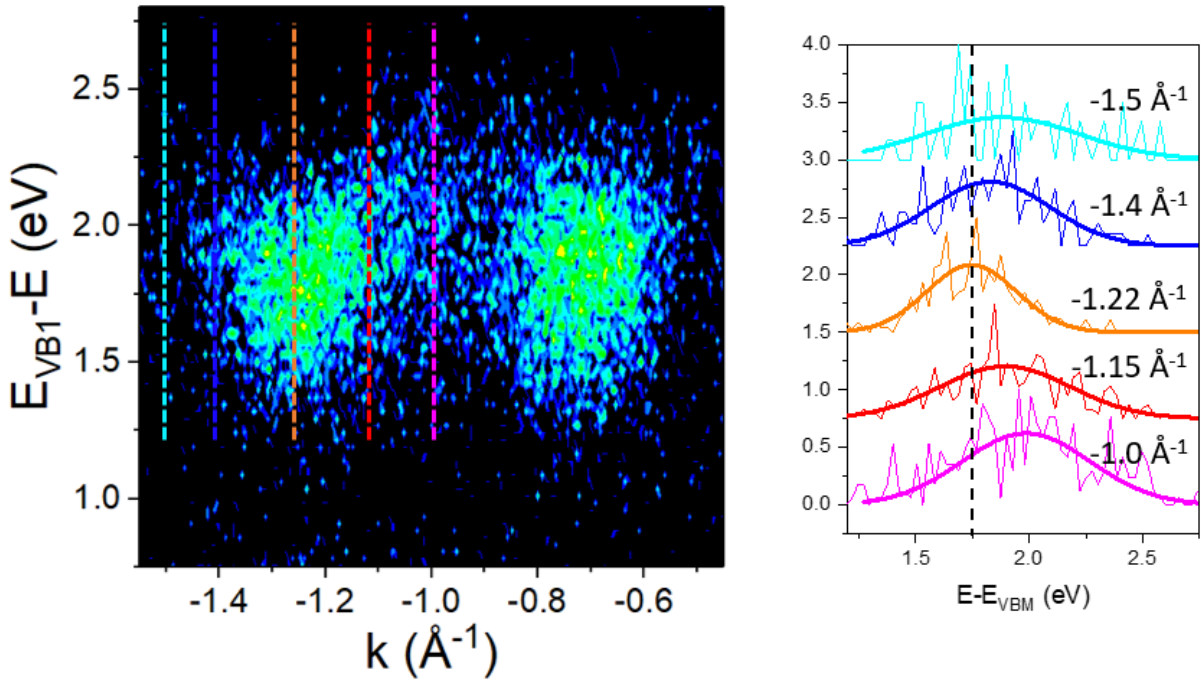

**Fig.S10:** (left) ARPES measurement at zero time delay for above gap excitation. The dashed color lines show the momentum cuts considered to plot the energy spectra. (right) Normalized energy spectra at different momentum of the K-valley from -1.5 to -1.0 Å$^{-1}$. In solid lines are represented the corresponding Gaussian fits.

9. Rise times for resonant excitation

In Fig. S11, we show a zoomed and normalized version of the data presented in Fig.3c clearly showing that there is delay of ~300fs in the exciton signals between the K-and Σ -valleys.

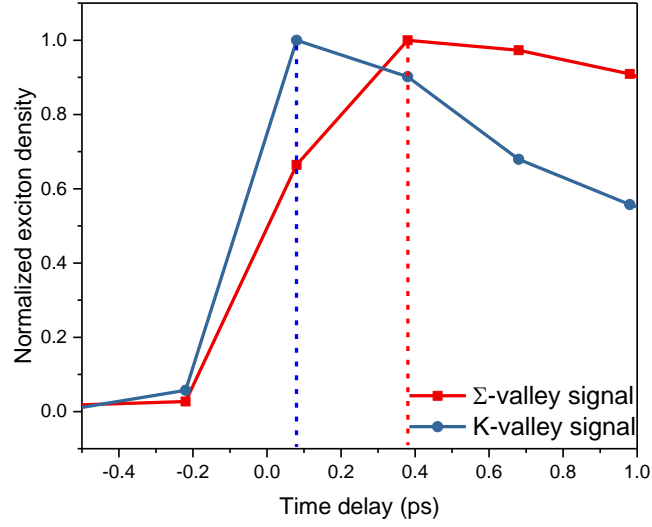

**Fig.S11:** Normalized exciton density at the K- (blue) and Σ- (red) valleys for a resonant excitation.

## 10. Momentum resolved holes and photoemission spectra at different time-delays

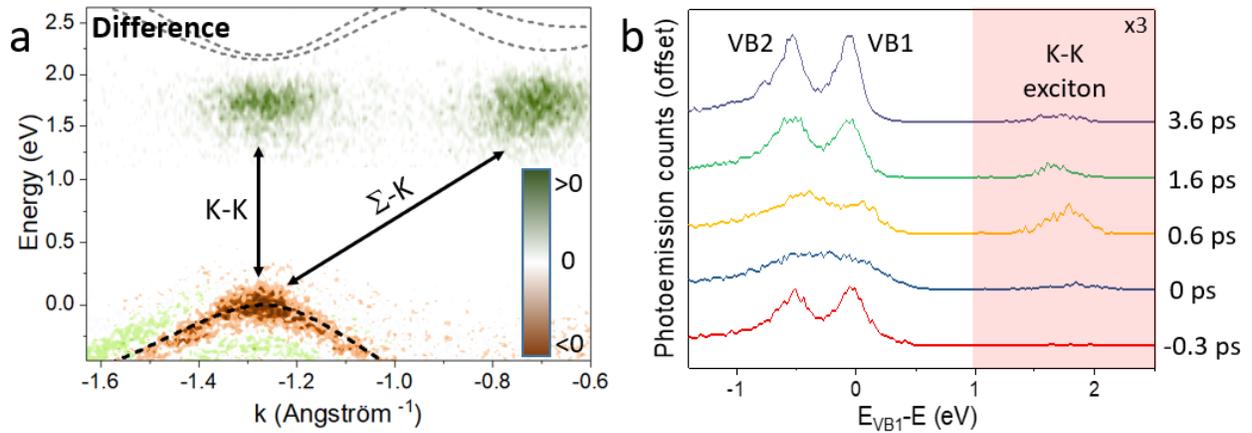

**Fig.S12:** (a) Difference image between no pump and after optical excitation showing a depletion of electron in the valence band associated to holes and photoemitted electron at the K- and Σ-valleys. We observe a broadening of the valence band as already reported by transient absorption measurements [44]. (b) Photoemission energy spectra at the K valley from -0.3 ps to 3.6 ps time delays showing the evolution of the valence bands and K-K exciton populations. A x3 multiplier is introduced from 1 eV to 2.5 eV for clarity.